# Quantifying the Excitonic Static Disorder in Organic Semiconductors

*Austin M. Kay,[1] Oskar J. Sandberg[1]\*, Nasim Zarrabi[1], Wei Li, Stefan Zeiske[1], Christina Kaiser[1], Paul Meredith, and Ardalan Armin[1]\**

Sustainable Advanced Materials (Sêr-SAM), Department of Physics, Swansea University, Singleton Park, Swansea SA2 8PP, United Kingdom

Email: ardalan.armin@swansea.ac.uk; o.j.sandberg@swansea.ac.uk

## Abstract

Organic semiconductors are disordered molecular solids and as a result, their internal charge dynamics and ultimately, the performance of the optoelectronic devices they constitute, are governed by energetic disorder. To ascertain how energetic disorder impacts charge generation, exciton transport, charge transport, and the performance of organic semiconductor devices, an accurate approach is first required to measure this critical parameter. In this work, we show that the static disorder has no relation with the so-called Urbach energy in organic semiconductors. Instead, it can be obtained from photovoltaic external quantum efficiency spectra at wavelengths near the absorption onset. We then present a detailed methodology, alongside a computational framework, for quantifying the static energetic disorder associated with singlet excitons. Moreover, the role of optical interference in this analysis is considered to achieve a high-accuracy quantification. Finally, the excitonic static disorder was quantified in several technologically-relevant donor-acceptor blends, including high-efficiency PM6:Y6.

# 1. Introduction

Organic semiconductors combine the electronic advantages of semiconducting materials, including the ability to conduct electricity and absorb or emit light, with the chemical and mechanical benefits of organic compounds such as flexibility, ease of processing, and tailorable optical properties.[1-3] Such properties make organic semiconductors suitable for several applications, including organic light-emitting diodes (OLEDs)—a key component of the thin-film display technology utilised in smartphones. Other avenues-of-research that have recently attracted considerable attention include the use of organic semiconductors in photovoltaic applications such as organic solar cells and indoor photovoltaics.[3-8] The efficiency at which organic solar cells convert solar energy into electrical power, the power conversion efficiency (PCE), has seen a steady increase over the past few years with 18% now in hand and the milestone of 20% expected to be reached imminently.[3, 9, 10] Currently, these cells suffer from a number of limitations including insufficient light trapping, short exciton diffusion lengths, large non-radiative recombination losses, and low charge carrier mobilities.[11] Provided that these limitations could be overcome, organic solar cells are anticipated to eventually reach a maximum PCE of around 25%.[3] Compared with lead-halide perovskite solar cells with PCEs greater than 25%, it is clear that further investigations must be conducted to push the PCE of organic solar cells to its predicted theoretical boundary, allowing them to be commercially viable.[12]

One of the first-order limiting factors on the PCE of an organic solar cell is the static energetic disorder of the active organic semiconductor layer, an inherent characteristic of molecular solids. The static disorder is defined by the broadness of the density of states (DOS), commonly approximated by either a Gaussian or an exponential distribution.[13] An increased static disorder generally correlates with an increased number of trap-like states in the gap and is typically associated with a degraded transport of excitations and charges through the active layer.[14-19] Nevertheless, the role of static disorder in organic solar cells has remained controversial. Increased levels of energetic disorder have been shown to result in reduced fill factors (FFs), short circuit current densities ($J_{SC}$), and open-circuit voltages ($V_{OC}$).[17, 20-22] Conversely, it has been suggested that higher levels of static disorder near the donor-acceptor (D:A) interface facilitates charge separation, thus enhancing charge generation and reducing recombination.[23] Therefore, while a low disorder for free charges across the bulk is desired to ensure efficient transport, an increased level of energetic disorder associated with charge-transfer (CT) states at the D:A interface might be beneficial for the charge generation-recombination dynamics.[24]

With the emergence of low-offset D:A systems based on non-fullerene acceptors (NFAs), the role of exciton dynamics has become increasingly important, defining the radiative PCE limit in state-of-the-art organic solar cells.[25-29] However, the effect of the excitonic disorder on the charge-recombination dynamics has remained poorly understood. By quantifying, understanding, and minimising the (potentially) adverse effects of excitonic disorder, the device performance could be optimised in organic solar cells, bringing the PCE closer to its theoretical limit. To this end, a reliable approach is therefore required to quantify the excitonic static disorder.

Energetic disorder in organic semiconductors is typically quantified using one of a few different methodologies. These include Kelvin probe and temperature-dependent charge carrier mobility measurements to probe the disorder of charges, and temperature-dependent photovoltaic external quantum efficiency ($EQE_{PV}$) measurements to estimate the static disorder of CT states.[30-34] A method for quantifying the energetic disorder of excitons, based on wavelength-dependent internal quantum efficiency measurements in neat organic semiconductors, was recently proposed by Hood et al..[35] While this method has several benefits, including single-temperature measurements, the technique cannot be adapted from neat organic semiconductors to organic D:A blends. In blends, the excitonic disorder has been inferred from the Urbach energy extracted from the sub-gap absorption tail, assuming an exponential shape and distribution.[36] However, the spectral range is often limited by the absorption of CT states and mid-gap trap states.[37, 38] Furthermore, Kaiser et al. recently found that the apparent Urbach energy in several technologically-relevant blend systems is strongly energy-dependent, is consistent with a Gaussian DOS rather than an exponential DOS, and saturates to the thermal energy at low enough photon energies.[29] Finally, as it is challenging to directly measure the sub-gap absorption coefficient, it must instead be inferred from the $EQE_{PV}$, the line-shape of which is affected by optical interference effects.[39, 40]

In this work, we propose a methodology (along with a computational framework) for quantifying the static energetic disorder associated with excitons from EQE$_{PV}$ spectra, allowing for excitonic static disorder of several systems to be estimated. For an improved accuracy, we show that the effects of optical interference on the EQE$_{PV}$ spectra must first be minimised. To do this, we propose the use of two transparent device architectures. The first architecture is a thin-film photovoltaic device with a photoresistor-like architecture, the 'lateral structure', that offers minimal optical interference effects at the expense of poor charge collection. The second architecture is an organic solar cell with a transparent top electrode that results in far less internal reflection (compared with the conventional organic solar cell architecture). Using minimal-interference devices, we have quantified the excitonic static disorder in five organic D:A blends, including a PM6:Y6 system.

## 2. Results and Discussion

### 2.1. Methodology for Quantifying Excitonic Disorder from Sub-Gap Absorption

In the work described here, we quantified the static energetic disorder associated with excitons, $\sigma_S$, by applying a model for the sub-gap singlet exciton (SE) absorption coefficient ($\alpha_{SE}$) to the spectral regime characterised by SE absorption. In D:A blends, $\alpha_{SE}$ is characterised by excitons in the component with the narrower gap; hence, the relevant optical gap ($E_{opt}$) can be expressed as $E_{opt} = \min(E_D, E_A)$, where $E_D$ and $E_A$ are the SE energies of the donor and acceptor, respectively. In the model, we assume the sub-gap absorption coefficient to be of the form $\alpha_{SE} = \int_0^\infty \alpha_0(E'_{SE}) g_{DOS}(E'_{SE}) dE'_{SE}$, where $\alpha_0(E'_{SE})$ is the absorption coefficient of a singlet exciton mode at energy $E'_{SE}$, while $g_{DOS}$ describes the excitonic density of states. In the case of a Gaussian distribution, this DOS is centred around the mean optical gap ($E_{opt}$), has a standard deviation $\sigma_S$ that quantifies the excitonic static disorder, and is given by

$$g_{DOS}(E'_{SE}) = \frac{N_{SE}}{\sqrt{2\pi\sigma_S^2}} \exp\left(-\frac{[E'_{SE} - E_{opt}]^2}{2\sigma_S^2}\right), \tag{1}$$

where $N_{SE}$ is the number density of excitonic states in the active layer.

The absorption of CT states of a single energy mode has generally been described as a Marcus charge-transfer process or with related extensions including molecular vibrations.[37, 41, 42] For excitons, however, Kaiser et al. recently suggested that sub-gap SE absorption is better described by a Boltzmann factor.[29] In that framework, the resulting absorption coefficient of an SE mode of energy $E'_{SE}$, at an incident photon energy $E$, can be approximated as

$$\alpha_0(E, E'_{SE}) \approx \alpha_{sat} \begin{cases} 1 & , \text{ for } E \geq E'_{SE}, \\ \exp\left[-\dfrac{E'_{SE} - E}{k_B T}\right] & , \text{ for } E < E'_{SE}, \end{cases} \quad (2)$$

where $k_B$ is the Boltzmann constant, $T$ is the temperature, and $\alpha_{sat}$ is a pre-factor that holds a weak photon energy-dependence. Such a dependence can also be justified in terms of a generalised non-adiabatic Marcus theory, assuming the electronic transfer to be between a localised ground state and a diffuse excited state (in the weak-coupling limit).[29, 43-45] The associated sub-gap absorption coefficient, determined by integrating the product of equations (1) and (2) with respect to $E'_{SE}$, is obtained as

$$\alpha_{SE}(E) = \frac{\alpha_{sat}}{2} \left\{ \exp\left[\frac{E - E_{opt} + \frac{\sigma_S^2}{2k_B T}}{k_B T}\right] \left(1 - \mathrm{erf}\left[\frac{E - E_{opt} + \frac{\sigma_S^2}{k_B T}}{\sigma_S \sqrt{2}}\right]\right) + 1 + \mathrm{erf}\left[\frac{E - E_{opt}}{\sigma_S \sqrt{2}}\right] \right\}, \quad (3)$$

where erf denotes the error function.[29]

In practice, it is challenging to directly measure the sub-gap absorption coefficient, $\alpha$, of an optoelectronic device.[39] Instead, in this work, we inferred the sub-gap absorption coefficient from sensitive $EQE_{PV}$ measurements, which are superior in sensitivity to other methods such as Fourier transform photocurrent and photothermal deflection spectroscopy.[46] In the case of weakly-absorbing sub-gap features, characterised by $\alpha(E)d_{AL} \ll 1$ (where $d_{AL}$ is the active layer thickness), it is usually assumed that $EQE_{PV}(E) \propto \alpha(E)d_{AL}$. However, this condition may fail for sub-gap excitons near the optical gap and therefore, to account for absorption in the active layer, we write

$$EQE_{PV}(E) \approx EQE_0 [1 - e^{-\alpha(E)d_{AL}}], \quad (4)$$

where $EQE_0$ is a pre-factor that, in general, holds an energy dependence.[40] However, in this work we assume $EQE_0$ to be constant within the spectral range of interest, corresponding to a uniform net intensity transmission into the active layer, minimal back-reflection, and a wavelength-independent internal quantum efficiency in the device.[47, 48] Hence, the static disorder can be quantified by applying equations **(3)** and **(4)** to the spectral regime characterised by $\alpha(E) = \alpha_{\text{SE}}(E)$.

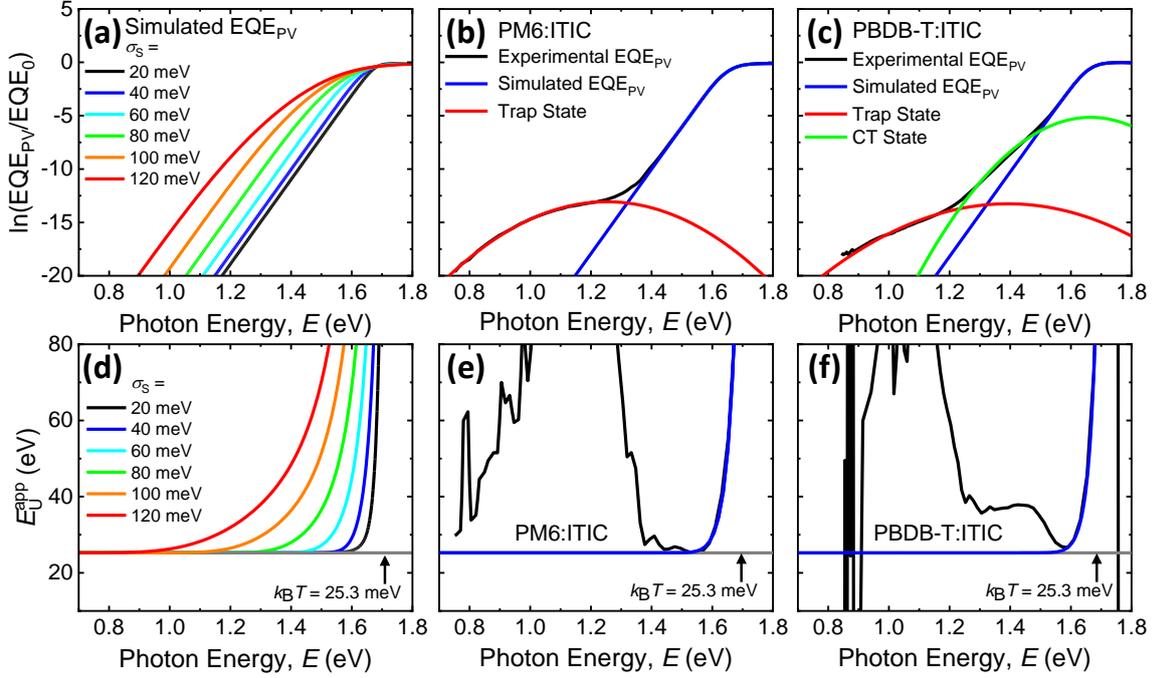

**Figure 1**. A comparison between theoretical and experimental $\text{EQE}_{\text{PV}}$ spectra and the corresponding apparent Urbach energy spectra. **(a)** The natural logarithm of sub-gap $\text{EQE}_{\text{PV}}$ tails characterised by singlet exciton absorption, created using equations **(3)** and **(4)**, with $E_{\text{opt}} = 1.60$ eV, $k_{\text{B}}T = 25.3$ meV, and $\sigma_{\text{S}}$ varied from 20 meV to 120 meV. **(b)** and **(c)** show the sensitive $\text{EQE}_{\text{PV}}$ spectra of PM6:ITIC and PBDB-T:ITIC bulk heterojunctions respectively. The Gaussian absorption features of trap states and charge-transfer (CT) states are indicated by the red and green lines, respectively. The apparent Urbach energy spectra shown in **(d)**-**(f)** were determined from **(a)**-**(c)** using equation **(5)**.

To precisely identify the excitonic sub-gap regime and to avoid other, more subtle absorption features present in organic systems, including trap state and CT state characteristics, we utilise the (energy-dependent) apparent Urbach energy ($E_{\text{U}}^{\text{app}}$), given by[29]

$$E_{\text{U}}^{\text{app}}(E) = \left[\frac{\text{d}\ln[\text{EQE}_{\text{PV}}(E)]}{\text{d}E}\right]^{-1}. \tag{5}$$

Through equation (**5**), the subtle absorption characteristics of CT states and trap states result in prominent features in the apparent Urbach energy and therefore, the spectral regime corresponding to SE absorption is identified more easily. **Figure 1a** shows sub-gap absorption tails obtained using equations (**3**) and (**4**), assuming $E_{\text{opt}} = 1.60$ eV and $k_B T = 25.3$ meV, for $\sigma_S$ varied from 20 meV to 100 meV. Based on equation (**3**), in the case of sub-gap SE absorption, two distinct regimes can be identified. At small photon energies, $\alpha_{\text{SE}}$ grows exponentially with $E_U^{\text{app}}(E) \to k_B T$, regardless of the magnitude of the excitonic static disorder, as shown by the coalescence of the simulated apparent Urbach energy spectra with the dashed, grey line in **Figure 1d**. In turn, at higher photon energies, the sub-gap absorption eventually takes a Gaussian-like shape as the apparent Urbach energy becomes strongly energy dependent, increasing by orders of magnitude.

Finally, to demonstrate the applicability of our model, we apply it to the experimentally-determined $EQE_{\text{PV}}$ spectra of PM6:ITIC and PBDB-T:ITIC bulk-heterojunctions (BHJs) with inverted device architectures, as shown in **Figure 1b** and **1c**. The corresponding apparent Urbach energy spectra are shown in **Figure 1e** and **1f**. Using these spectra, the points where the line shapes cease to be characterised by the behaviour of SEs were identified. Below these points, deviations due to the effects of trap states (and CT states in higher-offset systems like PBDB-T:ITIC) become apparent. Following this, equations (**3**) and (**4**) were applied to the appropriate regimes of **Figure 1c** and **1d**, with values for parameters-of-interest being extracted and used to simulate the SE-characterised $EQE_{\text{PV}}$ and $E_U^{\text{app}}$ spectra shown by the blue lines. From **Figure 1**, it is evident that, when a Gaussian DOS is assumed, the excitonic sub-gap behaviour can be described well by equation (**3**). It should be noted that this behaviour cannot be reproduced if an exponential DOS and/or Marcus charge-transfer is instead assumed in equation (**1**) and equation (**2**), respectively.

The methodology for accurately quantifying the excitonic static disorder is summarised in two steps: Firstly, the lower limit of the application of equations (**3**) and (**4**) to an $EQE_{\text{PV}}$ spectrum is identified using the apparent Urbach energy. This lower limit is specified as the point where the line-shape ceases to be SE-characterised, whether that be due to the effects of CT states, trap states, or optical interference. Secondly, equations (**3**) and (**4**) are not applied beyond the first saturation peak due to the weak energy-dependence of the pre-factor, $\alpha_{sat}$, playing a more pivotal role. Both of these criteria were applied when applying the SE absorption model as neglecting them would result in an inaccurate quantification of the excitonic static disorder.

## 2.2. Accounting for the Effects of Optical Interference

With the methodology for quantifying energetic disorder established, it was applied to the $EQE_{PV}$ spectra of several systems to quantify their excitonic static disorder. The $EQE_{PV}$ spectra measured for PM6:ITIC BHJs of varied active layer thickness, $d_{AL}$, are shown in **Figure 2a**. In addition, the values for the excitonic static disorder and the optical gap, extracted by fitting these curves with equations **(3)** and **(4)**, are plotted in **Figure 2b**.

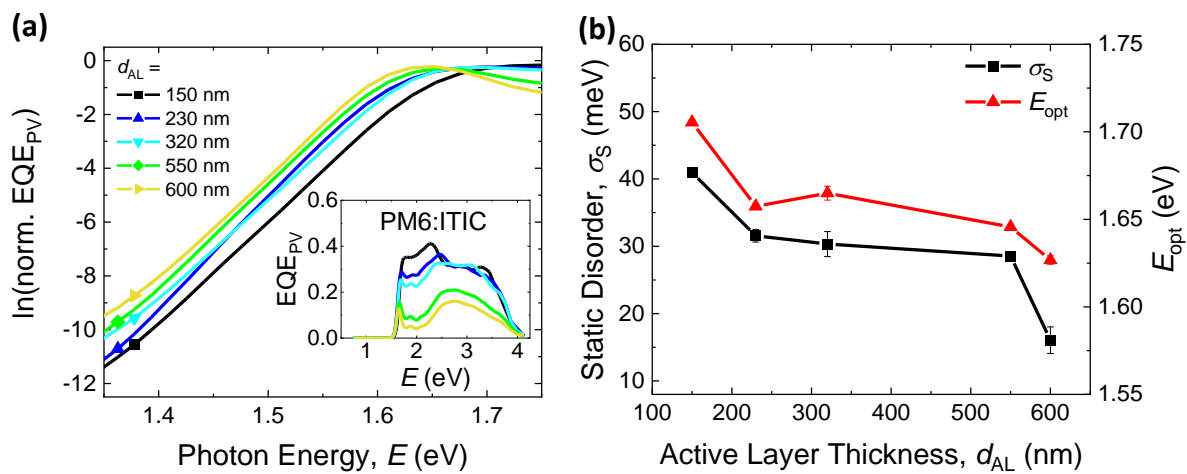

**Figure 2**. The apparent active layer thickness-dependence of the excitonic static disorder values extracted from $EQE_{PV}$ measurements for PM6:ITIC BHJs with an inverted device architecture. **(a)** The natural logarithm of normalised $EQE_{PV}$ spectra measured for PM6:ITIC BHJs of varied active layer thickness at room temperature. **(b)** The values for the excitonic static disorder ($\sigma_S$) and the optical gap ($E_{opt}$) extracted through the application of equations **(3)** and **(4)** to the normalised $EQE_{PV}$ spectra of **(a)**.

As shown in **Figure 2b**, an apparent active layer thickness-dependence can be seen in the excitonic disorder values extracted for PM6:ITIC BHJs. A similar dependence was also observed in several other systems, including PBDB-T:ITIC, PM6:BTP-eC9 and PM6:Y6 systems (see **Figure S1** in the Supplementary Information). Although slight variations could be expected in the static disorder values extracted for two BHJs of a given blend of differing active layer thickness (e.g., due to thickness-dependent variations in the fabrication techniques), it is unlikely that the excitonic static disorder would vary to the extent it appears to in **Figure 2b**. A possible origin behind these thickness-dependent variations in the extracted static disorder values could be subtle fluctuations in the line shapes of the sub-gap absorption tails, stemming from the optical interference effects that arise inside thin-film photovoltaics like BHJs. To explore these optical interference effects further, EQE$_{PV}$ spectra were simulated using an optical transfer-matrix model.[49] In this model, excitons are generated in the active layer of a multi-layered structure through the absorption of photons from an optical electric field distribution. This distribution is determined by accounting for the transmission and reflection of light at each interface using the refractive index, $\eta(E)$, and extinction coefficient, $\kappa(E)$, of all layers (for the optical constants of all layers bar the active layer, see **Figure S2** of the Supplementary Information). To determine the effects of optical interference using the optical transfer-matrix model, it was necessary to control the excitonic static disorder that characterised the active layer. Therefore, we employed two artificial sub-gap tails (shown in **Figure 3**) to describe the active layer's extinction coefficient in both the low-disorder and high-disorder limit, with one being characterised by $\sigma_S = 40$ meV and the other by $\sigma_S = 100$ meV. These artificial sub-gap extinction tails were created by first simulating sub-gap absorption tails using equation **(3)** with $\alpha_{sat} = 1.11$ nm$^{-1}$, $E_{opt} = 1.55$ eV, and $k_B T = 25.3$ meV. Following this, the absorption tails were converted to extinction tails using $\alpha(E) = 4\pi\kappa(E) / \lambda$, then appended to the experimental above-gap data.[49]

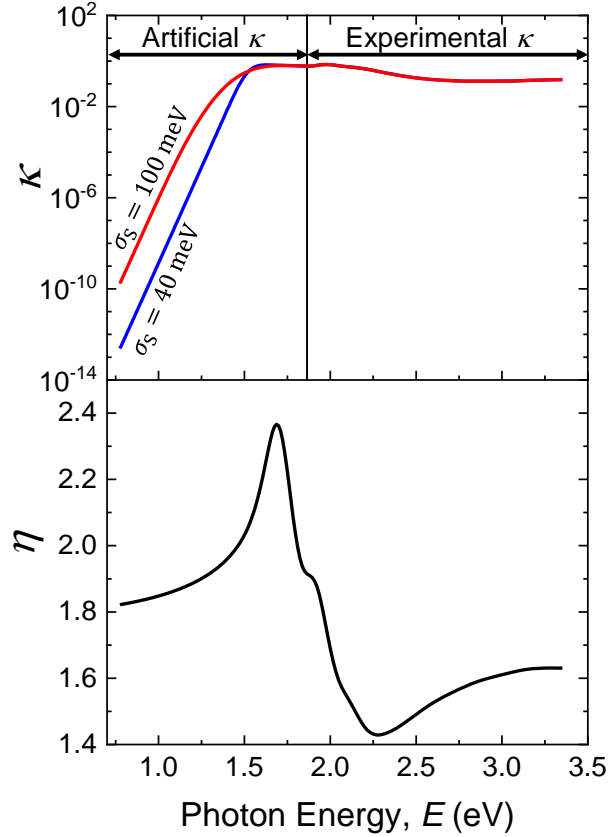

**Figure 3**. The optical constants used to describe the active layer in the optical transfer-matrix model, including artificial sub-gap extinction tails characterised solely by SE absorption. The sub-gap extinction tails were adapted from sub-gap absorption tails created using equation (**3**), with $\alpha_{\text{sat}} = 1.11$ nm$^{-1}$, $E_{\text{opt}} = 1.55$ eV, $k_B T = 25.3$ meV, and either $\sigma_S = 40$ meV or $\sigma_S = 100$ meV, then appended to the experimental above-gap data. The refractive index values were wholly experimental and describe a PM6:ITIC active layer.

Alongside the benefit of having two control values for the static disorder, with which extracted static disorder values could be compared, there were two other motives behind simulating the sub-gap tails: Firstly, the line shapes of the resultant sub-gap EQE$_{\text{PV}}$ tails would only be characterised by SE absorption and the effects of CT states and trap states would not be limit the suitable spectral regimes. Secondly, as optical interference effects were taken into account in the optical transfer-matrix model, conclusions could be drawn as to whether or not the apparent thickness-dependence was a result of the organic solar cells behaving as low-finesse optical cavities.

To explore and overcome the effects of optical interference effects, the methodology for quantifying the excitonic static disorder was applied to the simulated $EQE_{PV}$ spectra of devices of three hypothetical architectures. The first architecture, shown in **Figure 4a**, is that of an inverted organic solar cell with a highly-reflective top electrode (silver, Ag). In this figure, 'ITO' and 'ZnO' stand for the transparent conductor indium tin oxide and the electron transport layer zinc oxide, respectively. The second architecture, shown in **Figure 4b**, is very similar to the first but instead of a highly-reflective top electrode, it possesses a transparent one (indium zinc oxide, abbreviated to 'IZO') that produces far less internal-reflection and consequently, resulting in reduced optical interference in the active layer. Finally, the third architecture that was investigated (shown in **Figure 4c**) was an active layer film deposited onto a glass substrate—this architecture is herein referred to as the 'lateral structure'. Alongside the minimal number of interfaces that lateral structures possess, the use of small metallic electrodes at opposite ends of the device minimises the amount of back-reflection, culminating in negligible optical interference inside the active layer.

Using the optical transfer-matrix model, the $EQE_{PV}$ spectra shown in **Figure 4d-i** were simulated for all three device architectures using both sub-gap tails shown in **Figure 3**. In each case, the internal quantum efficiency for photon conversion was assumed to be spectrally-flat (and equal to unity for simplicity) and the effect of the incoherent glass layer at the forefront of each device was taken into account.[50, 51] For each device architecture, the thickness of the active layer ($d_{AL}$) was varied from 50 nm to 350 nm to explore any potential $d_{AL}$-dependencies in the extracted static disorder values. Comparing **Figure 4d** with **Figure 4e** and **4f**, the effect of optical interference on the line shapes of the sub-gap tails of the $EQE_{PV}$ spectra for organic solar cells with a reflective top electrode are immediately apparent. The sub-gap tails are quite disorganised compared with the line shapes of the lateral structure sub-gap tails. There is some uniformity across the sub-gap tails for the organic solar cells with transparent top electrodes, but not to the extent that is observed for the lateral structures. The same conclusion can be drawn from the $EQE_{PV}$ spectra simulated using $\sigma_S = 100$ meV, shown in **Figure 4g-i**.

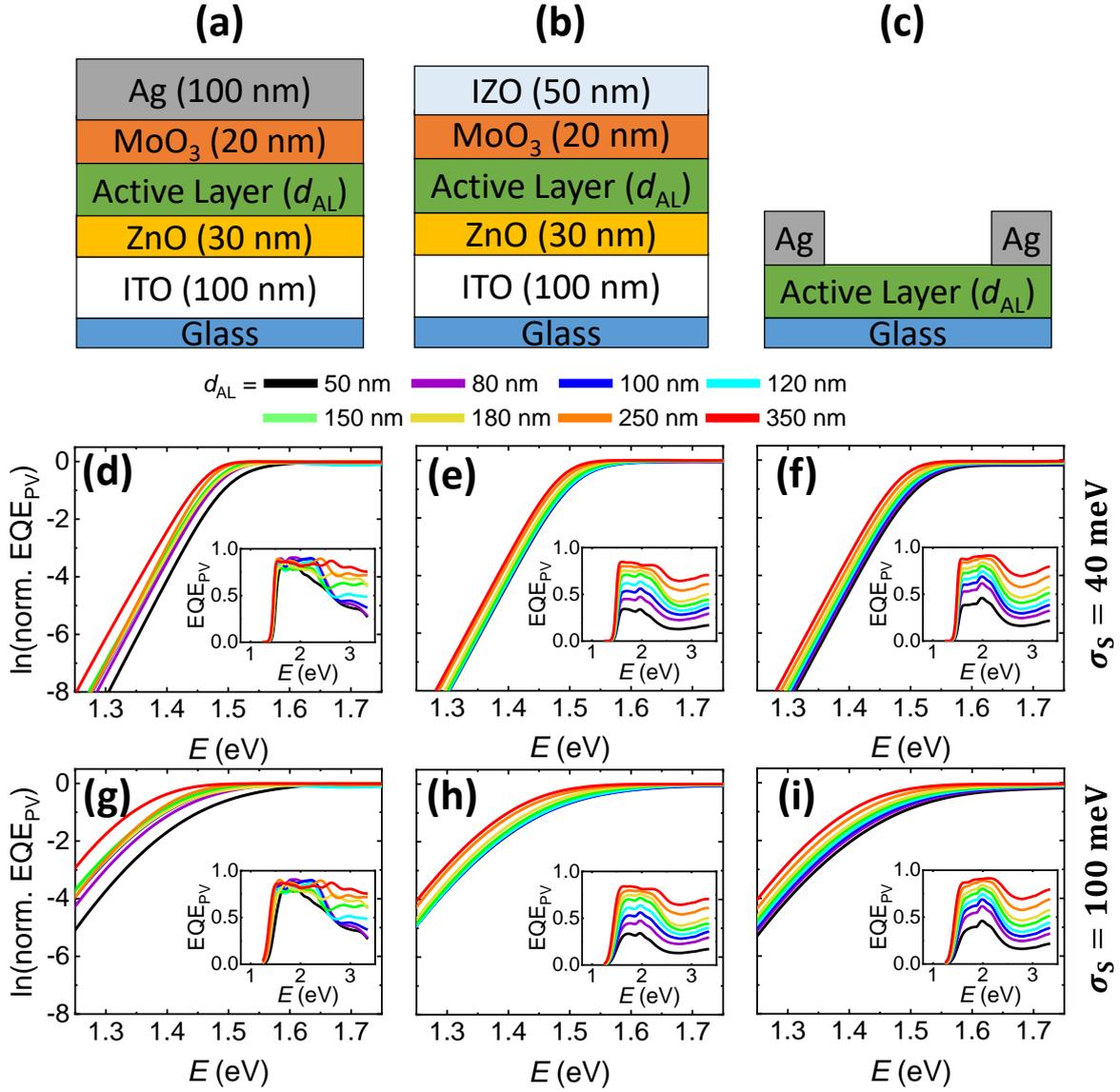

**Figure 4**. Simulating EQE$_{PV}$ spectra for devices of varied active layer thickness, for three architectures. **(a)-(c)** The three device architectures considered in these simulations: organic solar cells with **(a)** reflective top electrodes and **(b)** transparent top electrodes, and **(c)** lateral structures. The natural logarithm of the normalised EQE$_{PV}$ spectra, simulated for devices of each architecture with a variety of active layer thickness (from 50 nm to 350 nm), are shown for the $\sigma_S = 40$ meV case in **(d)-(f)** and for the $\sigma_S = 100$ meV case in **(g)-(i)**.

To minimise the error in the calculated static disorder values due to inappropriate choices in fitting ranges, a MATLAB script (provided in the Supplementary Information) was developed to automate the fitting process. The script applies equations (**3**) and (**4**) to the appropriate spectral regime of an input $EQE_{PV}$ spectrum and outputs values for the parameters-of-interest. For each of the $EQE_{PV}$ spectra shown in **Figure 4d-i**, the script conducted several unique fittings. For each spectrum, the best 20% of the fittings (in terms of adjusted R-square) were selected, and an average value for the excitonic static disorder was computed—the extracted values from all spectra are plotted as a function of active layer thickness in **Figure 5**. From this figure, it is clear that values extracted for the organic solar cells with a reflective top electrode fluctuate far more with increasing active layer thickness than the values extracted for both the organic solar cells with a transparent top electrode and the lateral structures.

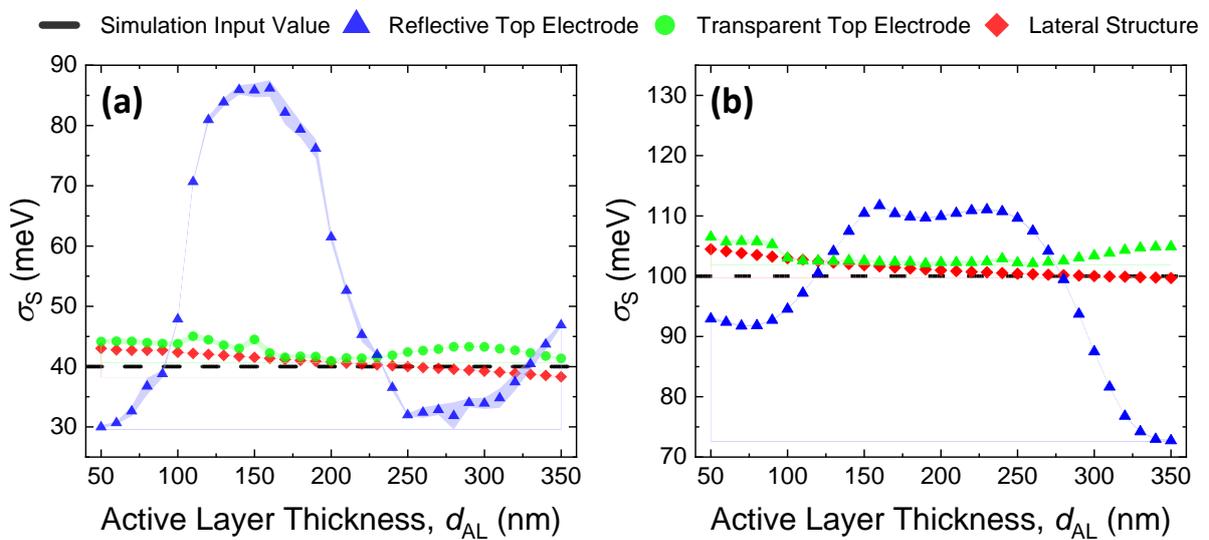

**Figure 5**. The excitonic static disorder values extracted from the simulated $EQE_{PV}$ spectra shown in **Figure 4d-i** using the MATLAB script provided in the Supplementary Information, for the three architectures illustrated in **Figure 4a-c** with active layers described by both **(a)** $\sigma_S = 40$ meV and **(b)** $\sigma_S = 100$ meV.

Three conclusions can be drawn from **Figure 5**: Firstly, the variations are very likely to be the result of optical interference, as demonstrated by, for example, the increasing absolute error in the excitonic disorder values obtained using organic solar cells with a reflective top electrode. The increasing deviations between the simulation values and the extracted values are likely due to variations in the pre-factor $EQE_0$ growing with increasing active layer thickness, leading to a larger degree of warping of the sub-gap tails and a more inaccurate quantification of the disorder.[40] Secondly, on average, the excitonic disorder is most accurately quantified in devices with a lateral structure, as shown by the close correlation between the input values ($\sigma_S = 40$ meV and $\sigma_S = 100$ meV) and the extracted values. Finally, the second-most accurate quantification of the static disorder is obtained using organic solar cells with a transparent top electrode. However, there are subtle oscillations with increasing active layer thickness. Nevertheless, for this architecture all extracted values are within around 5 meV of the simulation input values. Whereas the least-accurate quantification of the static disorder is obtained for the organic solar cells with highly-reflective metallic electrodes, indicating their impracticality for this application.

## 2.3. Experimental Results

To demonstrate the utility of devices with a lateral structure for quantifying disorder, the technique was applied to lateral structures of five different organic solar cell blends (PBDB-T:ITIC, PM6:O-IDTBR, PM6:Y6, PTB7-Th:CO*i*8DFIC, and PTB7-Th:PC$_{71}$BM). Sensitive EQE$_{PV}$ spectra were measured for each device (shown in **Figure S3** in the Supplementary Information) and the refined technique for quantifying the energetic disorder was applied—the extracted values are depicted in **Figure 6**. From this figure, it could be concluded that regardless of the uncertainty associated with the technique, it can be used to compare the energetic disorder associated with different blends. For example, the average excitonic static disorder of the PTB7-Th:CO*i*8DFIC active layer (63 meV) is almost twice the average disorder of the PBDB-T:ITIC active layer (34 meV).

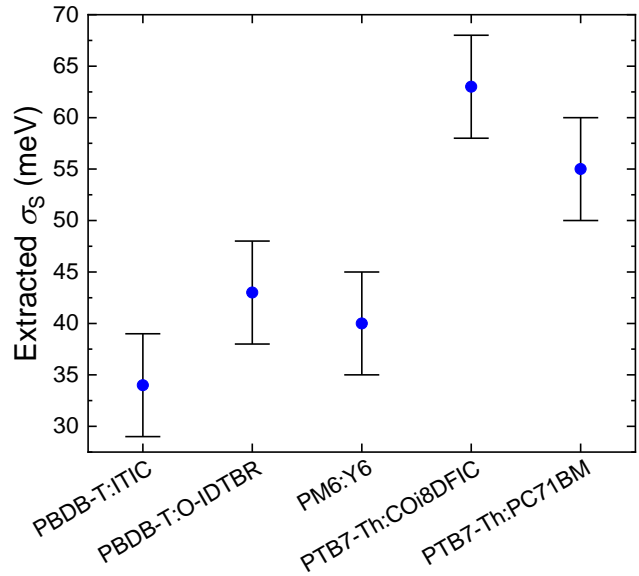

**Figure 6**. The excitonic static disorder values extracted for the lateral structures of a variety of organic solar cell blends, obtained through the application of equations **(3)** and **(4)** to the EQE$_{PV}$ spectra shown in **Figure S3** in the Supplementary Information. For each blend, a mean value for the excitonic static disorder was computed (where applicable) from the values shown in **Table S1** in the Supplementary Information.

## 3. Conclusion

Through the application of a model for the SE absorption coefficient, $\alpha_{SE}$, we have presented a detailed methodology and an associated computational tool for the determination of the excitonic static disorder, $\sigma_S$, in organic semiconductors. Utilising an optical transfer-matrix model, we have also investigated the adverse effects of optical interference on the methodology. To overcome the limitations presented by optical interference, we have proposed the use of two high-transparency device architectures and employed the lateral structure to quantify the energetic disorder of five systems-of-interest to the community. Using the methodology outlined in this work, alongside the MATLAB script that has been provided as supporting material, the static disorder can be quantified in several systems using room-temperature EQE$_{PV}$ measurements. Then, by comparing the excitonic static disorder in these systems with important parameters such as the exciton lifetime, or figures-of-merit such as the fill factor or the non-radiative $V_{OC}$ loss, any potential correlations between the two could be identified and investigated further.

## 4. Experimental Methods

### 4.1. Materials

**BTP-eC9:** 2,2'-[[12,13-Bis(2-butyloctyl)-12,13-dihydro-3,9-dinonylbisthieno[2'',3'':4',5']thieno[2',3':4,5]pyrrolo[3,2-e:2',3'-g][2,1,3]benzothiadiazole-2,10-diyl]bis[methylidyne(5,6-chloro-3-oxo-1H-indene-2,1(3H)-diylidene) ]]bis[propanedinitrile]

**CO*i*8DFIC:** 2,2′-[[4,4,11,11-tetrakis(4-hexylphenyl)-4,11-dihydrothieno[2′, 3′: 4,5]thieno[2,3-d]thieno[2′′′′, 3 ′′′′:4′′′,5′′′] thieno [2′′′,3′′′:4′,5′′]pyrano[2′′,3′′:4′,5'] thieno [2,3:4,5]thieno[3,2-b]pyran-2,9-diyl]bis[methylidyne(5,6-difluoro)]]

**ITIC:** 3,9-bis(2-methylene-(3-(1,1-dicyanomethylene)-indanone))-5,5,11,11-tetrakis(4-hexylphenyl)-dithieno[2,3-*d*:2′,3′-*d*′]-s-indaceno[1,2-*b*:5,6-*b*′]dithiophene

**O-IDTBR:** (5Z,5'Z)-5,5'-((7,7'-(4,4,9,9-tetraoctyl-4,9-dihydro-s-indaceno[1,2-b:5,6-b']dithiophene-2,7-diyl)bis(benzo[c][1,2,5]thiadiazole-7,4-diyl))bis(methanylylidene))bis(3-ethyl-2-thioxothiazolidin-4-one)

**PBDB-T:** Poly[(2,6-(4,8-bis(5-(2-ethylhexyl)thiophen-2-yl)-benzo[1,2-*b*:4,5-*b*′]dithiophene))-*alt*-(5,5-(1',3'-di-2-thienyl-5′,7′-bis(2-ethylhexyl)benzo[1′,2′-*c*:4′,5′-*c*′]dithiophene-4,8-dione)]

**PC$_{71}$BM:** [6,6]-phenyl-C71-butyric acid methyl ester

**PDINO:** 2,9-bis[3-(dimethyloxidoamino)propyl]anthra[2,1,9-def:6,5,10-*d'e'f'*]diisoquinoline-1,3,8,10(2*H*,9*H*)-tetrone

**PEDOT:PSS:** Poly(3,4-ethylenedioxythiophene) polystyrene sulfonate

**PM6:** Poly[(2,6-(4,8-bis(5-(2-ethylhexyl-3-fluoro)thiophen-2-yl)-benzo[1,2-*b*:4,5-*b*′]dithiophene))-*alt*-(5,5-(1′,3′-di-2-thienyl-5′,7′-bis(2-ethylhexyl)benzo[1′,2′-*c*:4′,5′-*c*′]dithiophene-4,8-dione)]

**PTB7-Th:** Poly[4,8-bis(5-(2-ethylhexyl)thiophen-2-yl)benzo[1,2-b;4,5-*b*']dithiophene-2,6-diyl-alt-(4-(2-ethylhexyl)-3-fluorothieno[3,4-b]thiophene-)-2-carboxylate-2-6-diyl)]

**Y6:** 2,2′-((2Z,2′Z)-((12,13-bis(2-ethylhexyl)-3,9-diundecyl-12,13-dihydro-[1,2,5]thiadiazolo[3,4-*e*]thieno[2′′,3′′:4′,5′]thieno[2′,3′:4,5]pyrrolo[3,2-*g*]thieno[2′,3′:4,5]thieno[3,2-*b*]indole-2,10-diyl)bis(methanylylidene))bis(5,6-difluoro-3-oxo-2,3-dihydro-1*H*-indene-2,1-diylidene))dimalononitrile

PC$_{71}$BM was purchased from Ossila (UK). O-IDTBR was purchased from Sigma-Aldrich (USA). CO*i*8DFIC, ITIC, PBDB-T, and PTB7-Th were purchased from Zhi-yan (Nanjing). BTP-eC9, PDINO, PM6, and Y6 were purchased from Solarmer (Beijing). PEDOT:PSS was purchased from Heraeus (Germany).

### 4.2. Device Fabrication

**Substrate preparation:** Commercially-patterned ITO coated glass and ultra-flat quartz-coated glass from Ossila were used for all devices in this work. All the substrates were sonicated in deionised water, acetone, and 2-propanol for 10 minutes each. The cleaned substrates were first dried by nitrogen and 110°C hotplate and then treated in UV-Ozone cleaner (Ossila, L2002A2-UK) for 20 minutes.

*4.2.1. Devices with an Inverted/Conventional Organic Solar Cell Architecture*

The thicknesses of all the following films were determined using ellipsometry.

**PBDB-T:ITIC** and **PM6:ITIC** thickness-dependent devices were fabricated with an inverted architecture (ITO/ZnO/Active Layer/MoO$_3$/Ag). The ZnO electron transport layer was prepared by dissolving 200 mg zinc acetate dihydrate in 2-methoxyethanol (2 mL) using ethanolamine (56 µL) as the stabilizer. The solution was stirred overnight under ambient conditions and spin-coated (4000 rpm for 30s) onto the ITO substrates then annealed at 200°C for 1 hour to obtain a thickness of approximately 30 nm. To fabricate devices with a PBDB-T:ITIC active layer, PBDB-T:ITIC with a D:A ratio of 1:1 was dissolved in a CB:DIO (99:1, v/v) solution to give various total concentrations. On the other hand, to fabricate devices with a PM6:ITIC active layer, PM6:ITIC with a D:A ratio of 1:1 was dissolved in a CF:DIO (99:1, v/v) solution to give various total concentrations. The exact concentrations and spin-coating speeds required to form the PBDB-T:ITIC and PM6:ITIC active layers (of varied thickness) considered in this work are given in **Table S2** the Supplementary Information. The as-cast films were then thermally annealed at 100°C for 10 minutes. Following this, 10 nm of MoO$_3$ and 100 nm of Ag was deposited on the active layer to form a cathode.

**PM6:BTP-eC9** and **PM6:Y6** thickness-dependent devices were fabricated with a conventional architecture (ITO/PEDOT:PSS/Active Layer/PDINO/Ag). The PEDOT: PSS solution was first diluted with the same volume of water, then cast at 4000 rpm onto the ITO substrate, followed by thermal annealing at 155°C for 15 minutes to produce a 10 nm film. To fabricate devices with a PM6:BTP-eC9 active layer, PM6:BTP-eC9 with a D:A ratio of 1:1.2 by weight was dissolved in a CF:DIO (99.5:0.5) solution to give various total concentrations. On the other hand, to fabricate devices with a PM6:Y6 active layer, PM6:Y6 with a D:A ratio of 1:1.2 by weight was dissolved in a CF:CN (99.5:0.5) solution to give a variety of total concentrations. The exact concentrations and spin-coating speeds required to form the PM6:BTP-eC9 and PM6:Y6 active layers (of varied thickness) considered in this work are given in **Table S3** of Supplementary Information. The active layers were further thermally annealed at 100ºC for 10 minutes. Following this, a PDINO solution of concentration 1 mg mL$^{-1}$ was spin-coated onto the active layers at 2000 rpm to form 10 nm films, and 100 nm of Ag was evaporated as to form the top electrode.

*4.2.2. Devices with a Lateral Structure*

Different to the traditional organic solar cell architecture, devices with a lateral structure were fabricated to avoid the optical interference effects that arise inside conventionally-structured OSCs. Devices with a lateral structure were fabricated by spin-coating the active layer onto a pre-cleaned glass substrate before evaporating 100 nm Ag to form a cathode and an anode separated by a distance of 30 µm. The exact D:A ratios, solvents, concentrations, and spin-coating speeds used to fabricate the PBDB-T:ITIC, PBDB-T:O-IDTBR, PTB7-Th:COi8DFIC, PTB7-Th:PC$_{71}$BM, and PM6:Y6 lateral structures are given in **Table S4** of the Supplementary Information.

### 4.3. Device Characterisation

**Photovoltaic external quantum efficiency (EQE$_{PV}$):** Sensitive photovoltaic external quantum efficiency (EQE$_{PV}$) measurements were conducted using a high-performance spectrophotometer (PerkinElmer, Lambda950) as a light source. The probe light beam was physically chopped at 273 Hz (Thorlabs, MC2000B) prior focusing on the device under test (DUT). The DUT photocurrent signal was measured by a lock-in amplifier (Stanford Research, SR860) in combination with a current pre-amplifier with an integrated, low-noise voltage source (FEMTO, DHPCA-200). For DUTs with conventional (lateral) device architecture, a bias voltage of 0 V (10 V) was applied. A detailed description of the EQE$_{PV}$ apparatus is provided elsewhere.[46]

## Acknowledgements

This work was funded through the Welsh Government's Sêr Cymru II Program 'Sustainable Advanced Materials' (Welsh European Funding Office − European Regional Development Fund). P.M. is a Sêr

Cymru II Research Chair and A.A. is a Rising Star Fellow also funded through the Welsh Government's Sêr Cymru II 'Sustainable Advanced Materials' Program (European Regional Development Fund, Welsh European Funding Office and Swansea University Strategic Initiative). This work was also funded by UKRI through the EPSRC Program Grant EP/T028511/1 Application Targeted and Integrated Photovoltaics.

## Conflicts of Interest

The authors declare no conflicts of interest.

# Supplementary Information

## Quantifying the Excitonic Static Disorder in Organic Semiconductors

*Austin M. Kay,[1] Oskar J. Sandberg[1]\*, Nasim Zarrabi[1], Wei Li, Stefan Zeiske[1], Christina Kaiser[1], Paul Meredith, and Ardalan Armin[1]\**

Sustainable Advanced Materials (Sêr-SAM), Department of Physics, Swansea University, Singleton Park, Swansea SA2 8PP, United Kingdom

Email: ardalan.armin@swansea.ac.uk; o.j.sandberg@swansea.ac.uk

**Table of Contents**

# 1. The Optical Interference Problem

## 1.1. Variation in Extracted Static Disorder Values

Once the technique for quantifying the excitonic static disorder was established, it was applied to the experimentally-determined EQE$_{PV}$ spectra of four systems where the devices had a variety of active layer thicknesses, $d_{AL}$. In **Figure 2** of the main text, it was shown that the excitonic static disorder values extracted from the sensitive EQE$_{PV}$ spectra of PM6:ITIC bulk heterojunctions (BHJs) had an apparent active layer thickness-dependence. As shown in **Figure S1**, this dependence was also observed in other systems, including PBDB-T:ITIC, PM6:BTP-eC9, and PM6:Y6. The sensitive EQE$_{PV}$ spectra measured for these devices are shown in **Figure S1a-c**, with the static disorder values extracted from fittings to the appropriate regimes being plotted as a function of active layer thickness in **Figure S1d**.

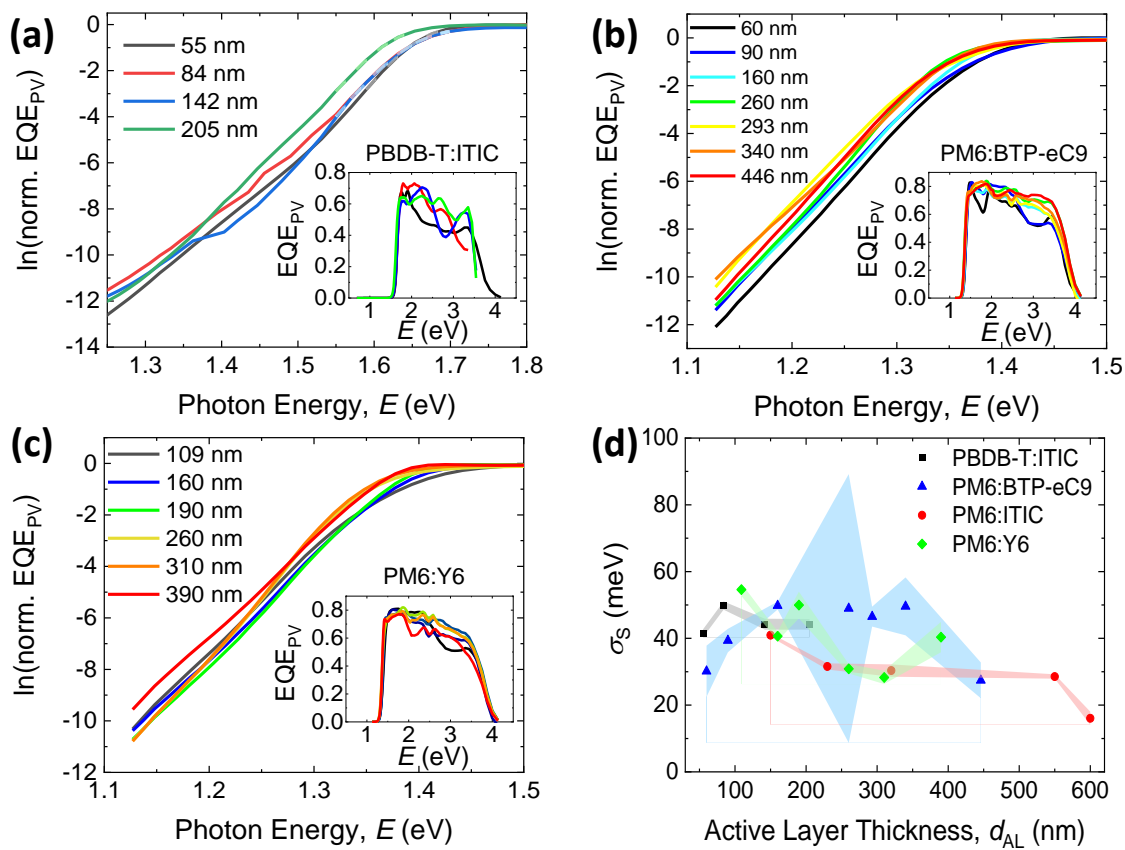

**Figure S1**. The apparent active layer thickness-dependence of the excitonic static disorder values extracted from the EQE$_{PV}$ spectra of four systems. **(a)-(c)** The natural logarithm of the normalised,

sensitive EQE$_{PV}$ spectra for PBDB-T:ITIC, PM6:BTP-eC9, and PM6:Y6 BHJs of a variety of active layer thicknesses. The excitonic static disorder values extracted from these spectra are plotted in **(d)**, where the shaded region indicated the statistical uncertainty obtained through the fittings.

## 1.2. Optical Modelling in Various Architectures

To investigate the effects of optical interference, and to test the benefits of using the transparent device architectures described in the main text, EQE$_{PV}$ spectra were simulated using an optical transfer-matrix model.[1, 2] In this model, the propagation of light through (and the generation of excitons in) the active layer of an organic solar cell is determined using the refractive index, $\eta(E)$, and extinction coefficient, $\kappa(E)$, of each layer of the cell. The experimentally-determined refractive index and extinction coefficient values used to describe the glass, ITO, IZO, ZnO, MoO$_3$, and Ag layers are shown in **Figure S2**. These values were utilised in combination with the refractive index and extinction coefficient values shown in **Figure 3** in the main text to simulate the EQE$_{PV}$ spectra shown in **Figure 4d-i**.

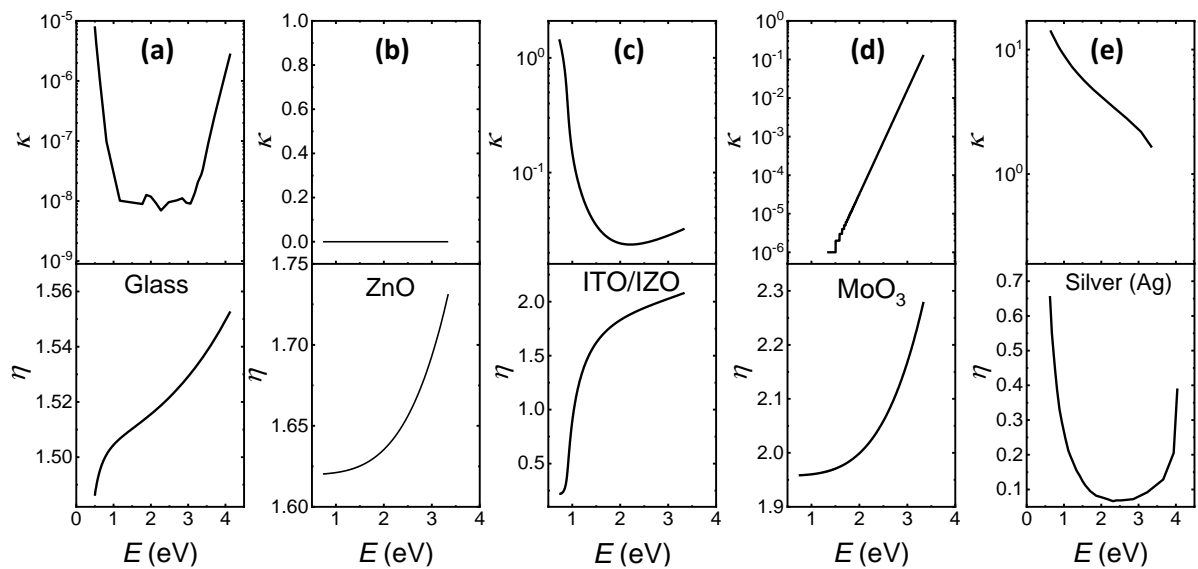

**Figure S2**. The real refractive index, $\eta(E)$, and extinction coefficient, $\kappa(E)$, values employed by the optical transfer-matrix model to describe **(a)** glass, **(b)** ZnO, **(c)** ITO and IZO (the same optical constants were used to describe both layers), **(d)** MoO$_3$, and **(e)** silver (Ag).

## 2. Lateral Structure Investigation

Having demonstrated that the excitonic static disorder is most accurately quantified for devices with a lateral structure, the energetic disorder of five technologically-relevant donor/acceptor blends was quantified. The fitted EQE$_{PV}$ spectra for each system (PBDB-T:ITIC, PBDB-T:O-IDTBR, PM6:Y6, PTB7-Th:CO$i$8DFIC, and PTB7-Th:PC$_{71}$BM) are shown in **Figure S3** and the extracted excitonic static disorder values are shown in **Figure 6** in the main text. Note that, in **Figure S3d**, the parameter '*D*' is the separation between the electrodes of the PTB7-Th:CO$i$8DFIC lateral structures. Where applicable, the static disorder values shown for the five blends in **Figure 6** are the averages across multiple devices. Moreover, the value given in **Table S1** for the PTB7-Th:CO$i$8DFIC lateral structure with $d_{AL} = 90$ nm is the average of the values extracted from the four EQE$_{PV}$ spectra of **Figure S3d**, corresponding to devices with $d_{AL} = 90$ nm and $D$ varied from 40 μm to 80 μm. For all other devices, the electrode separation was $D = 30$ μm.

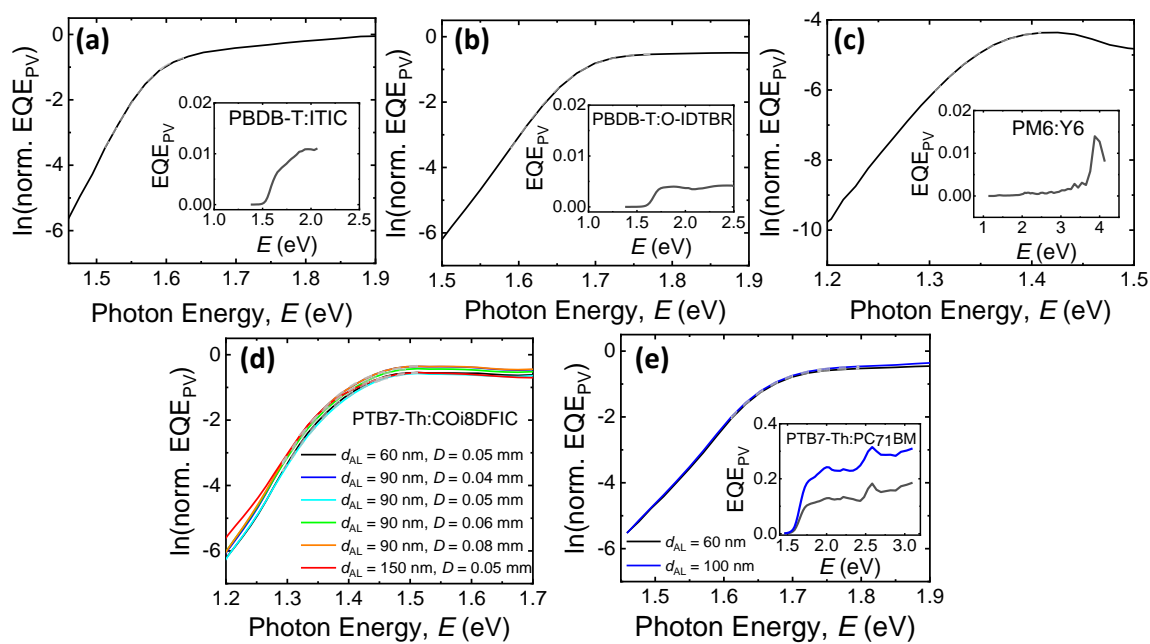

**Figure S3**. Experimental EQE$_{PV}$ spectra for systems with a lateral structure of varied active layer thicknesses. **(a)** The EQE$_{PV}$ spectrum for a PBDB-T:ITIC lateral structure with $d_{AL} = 90$ nm. **(b)** The EQE$_{PV}$ spectrum for a PBDB-T:O-IDTBR lateral structure with $d_{AL} = 200$ nm. **(c)** The EQE$_{PV}$ spectrum for a PM6:Y6 lateral structure with $d_{AL} = 100$ nm. **(d)** The EQE$_{PV}$ spectra for a PTB7-

Th:CO$i$8DFIC lateral structures with $d_{AL}$ varied from 60 nm to 150 nm. **(e)** The EQE$_{PV}$ spectra for a PTB7-Th:PC$_{71}$BM lateral structures with $d_{AL}$ = 60 nm and $d_{AL}$ = 100 nm.

**Table S1**. Excitonic static disorder values extracted through the application of equations **(3)** and **(4)** to the EQE$_{PV}$ spectra of **Figure S3**, automated using the provided MATLAB script.

| Blend | Active Layer Thickness, $d_{AL}$ (nm) | Extracted $\sigma_S$ Value (±5 meV) |
|---|---|---|
| PBDB-T:ITIC | 100 | 34 |
| PBDB-T:O-IDTBR | 200 | 43 |
| PM6:Y6 | 250 | 40 |
| PTB7-Th:CO$i$8DFIC | 60 | 65 |
| PTB7-Th:CO$i$8DFIC | 90 | 67 |
| PTB7-Th:CO$i$8DFIC | 150 | 57 |
| PTB7-Th:PC$_{71}$BM | 60 | 54 |
| PTB7-Th:PC$_{71}$BM | 100 | 56 |

## 3. Fabrication Techniques

**Table S2**. The solvents, concentrations, and spin-coating speeds used in this work to fabricate the PBDB-T:ITIC and PM6:ITIC active layers of the devices with an inverted architecture.

| Blend | Solvent | Active Layer Thickness, $d_{AL}$ (nm) | Solution Concentration (mg mL$^{-1}$) | Spin-Coating Speed (rpm) |
|---|---|---|---|---|
| PBDB-T:ITIC (1:1) | CB:DIO (99:1, v/v) | 55 | 18 | 2500 |
| | | 84 | 18 | 1500 |
| | | 142 | 18 | 700 |
| | | 205 | 22 | 900 |
| PM6:ITIC (1:1) | CF:DIO (99:1, v/v) | 150 | 18 | 5000 |
| | | 230 | 25 | 5000 |
| | | 320 | 35 | 5000 |
| | | 550 | 45 | 5000 |
| | | 600 | 50 | 5000 |

**Table S3**. The solvents, concentrations, and spin-coating speeds used in this work to fabricate the PM6:BTP-eC9 and PM6:Y6 active layers of the devices with a conventional architecture.

| Blend | Solvent | Active Layer Thickness, $d_{AL}$ (nm) | Solution Concentration (mg mL$^{-1}$) | Spin-Coating Speed (rpm) |
|---|---|---|---|---|
| PM6:BTP-eC9 (1:1.2) | CF:DIO (99.5:0.5) | 60 | 12 | 3000 |
| | | 90 | 16 | 3000 |
| | | 160 | 20 | 2000 |
| | | 260 | 25 | 2000 |
| | | 293 | 30 | 2000 |
| | | 340 | 35 | 2000 |
| | | 446 | 40 | 2000 |
| PM6:Y6 (1:1.2) | CF:CN (99.5:0.5) | 109 | 16 | 2000 |
| | | 160 | 20 | 4000 |
| | | 190 | 25 | 4000 |
| | | 260 | 30 | 4000 |
| | | 310 | 35 | 4000 |
| | | 390 | 40 | 4000 |

**Table S4.** The solvents, concentrations, and spin-coating speeds used in this work to fabricate the PBDB-T:ITIC, PBDB-T:O-TDBR, PM6:Y6, PTB7-Th:C and PM6:Y6 active layers of the devices with a lateral structure.

| Blend | Solvent | Active Layer Thickness, $d_{AL}$ (nm) | Solution Concentration (mg mL$^{-1}$) | Spin-Coating Speed (rpm) |
|---|---|---|---|---|
| PBDB-T:ITIC (1:1) | CB:DIO (99:1, v/v) | 100 | 16 | 1000 |
| PBDB-T:O-ITDBR (1:1.5) | CB:DIO (99:1, v/v) | 200 | 20 | 1000 |
| PM6:Y6 (1:1.2) | CF: 1-chloronaphthalene (99.5:0.5, v/v) | 250 | 25 | 2000 |
| PTB7-Th:CO$i$8DFIC (1:1.5) | CF:DIO (99:1, v/v) | 60 | 16 | 4000 |
| | | 90 | 16 | 2000 |
| | | 150 | 16 | 1000 |
| PTB7-Th:PC$_{71}$BM (1:1.5) | CB:DIO (97:3, v/v) | 60 | 18 | 2000 |
| | | 100 | 18 | 1100 |

## 4. The MATLAB Script

To automate the application of equations **(3)** and **(4)** to both simulated and experimentally-determined $EQE_{PV}$ spectra, a MATLAB script was prepared and utilised in this work. Using an estimate for the upper and lower limit of the spectral range-of-interest, the MATLAB script fits all unique combinations of minimum and maximum photon energies in this range (whilst ensuring a minimum of seven data points to maintain statistical significance). Values for $\alpha_{sat}$, $E_{opt}$, and $\sigma_S$ are extracted from each fitting, alongside goodness-of-fit parameters such as the adjusted R-squared and the root-mean-square-error. The extracted values for the excitonic static disorder are then organised in terms of their adjusted R-square values. Following this, the static disorder values corresponding to top 20% of fittings with adjusted R-square values closest to unity are sampled, indicating that the SE absorption model best describes these fittings, and it is unlikely that other absorption characteristics have an effect on them. The average value of the parameters-of-interest are then computed, including the static disorder with an uncertainty of ±5 meV.